# The Pioneer maser signal anomaly:
## Possible confirmation of spontaneous photon blueshifting

P. A. LaViolette
April 2005
*Physics Essays*, volume 18(2), 2005

*The Starburst Foundation, 1176 Hedgewood Lane, Niskayuna, NY 12309*
Electronic address: gravitics1@aol.com

**Abstract**
The novel physics methodology of subquantum kinetics predicted in 1980 that photons should blueshift their frequency at a rate that varies directly with negative gravitational potential, the rate of blueshifting for photons traveling between Earth and Jupiter having been estimated to average approximately $1.3 \pm 0.65 \times 10^{-18}$ s$^{-1}$, or $1.1 \pm 0.6 \times 10^{-18}$ s$^{-1}$ for signals traveling a roundtrip distance of 65 AU through the outer solar system. A proposal was made in 1980 to test this blueshifting effect by transponding a maser signal over a 10 AU round-trip distance between two spacecraft. This blueshift prediction has more recently been corroborated by observations of maser signals transponded to the Pioneer 10 spacecraft. These measurements indicate a frequency shifting of approximately $2.28 \pm 0.4 \times 10^{-18}$ s$^{-1}$ which lies within $2\sigma$ of the subquantum kinetics prediction and which cannot be accounted for in terms of known forces acting on the craft. This blueshifting phenomenon implies the existence of a new source of energy which is able to account for the luminosities of red dwarf and brown dwarf stars and planets, and their observed sharing of a common mass-luminosity relation.

**Résumé**
La nouvelle théorie de la physique du cinétique subquantique avait prévu en 1985 que les photons devraient se déplacer vers le bleu à un rythme qui varie directement avec le potentiel de la gravité négative. Le taux de déplacement de photons vers le bleu entre la Terre et Jupiter est estimé être approximativement $(1.3 \pm 0.65) \times 10^{-18}$ s$^{-1}$, ou de $(1.1 \pm 0.6) \times 10^{-18}$ s$^{-1}$ pour le chemin aller-retour de signaux d'une distance de 65 AU à travers la région extérieure du système solaire. Une proposition a été faite en 1980 afin de vérifier l'effet du déplacement vers le bleu par un signal maser de transpondeur sur une distance aller-retour entre deux engins spatiaux. Cette prédiction fut corroborée par l'observation des signaux maser de transpondeur vers le vaisseau spatial du Pionnier 10. Ces mesures indiquent un décalage de fréquence approximativement de $2.28 \pm 0.4 \times 10^{-18}$ s$^{-1}$, qui se trouve en dessous de $2\sigma$ de la prévision de la cinétique subquantique et qui n'est pas expliqué en terme des forces connues agissant sur le vaisseau. Ce phénomène de déplacement vers le bleu suggère l'existence d'une nouvelle source d'énergie qui explique les luminositiés des étoiles naines rouges et naines brunes, des planètes, et leur partage d'une relation de luminosité masse commune.

___

# 1. INTRODUCTION

During the past 30 years, 2.1 GHz maser signals have been transmitted to the Pioneer 10 and Pioneer 11 spacecraft and coherently transponded back to Earth, the frequency shift of the received signal being used to determine the recessional velocity of the spacecraft for purposes of navigation. However, Anderson *et al*.[1, 2] report that when the computed velocity is compared to the velocity predicted by orbital models, a discrepancy is found, even after adjustments are made for all known forces that might act on the spacecraft. They find a frequency blueshift residual that increases linearly with time, or in direct proportion to the increase in the line-of-sight distance to the spacecraft.

If interpreted as a Doppler effect, this residual implies the presence of an anomalous force accelerating the craft toward the Sun which Anderson *et al*. calculate to be $8.7 \pm 1.3 \times 10^{-8}$ cm/s$^2$. Although, as is discussed below, when the propulsive effects of on-board thermal radiation sources are taken into account, this decreases to a residual acceleration of $6.85 \pm 1.3 \times 10^{-8}$ cm/s$^2$. If interpreted as an anomalous acceleration, the effect is perplexing since most plausible forces, such as gravity, decrease rapidly with distance whereas the Pioneer apparent acceleration remains relatively constant with time. Moreover, an anomalous acceleration of similar magnitude does not appear to be acting on the planets, given that their orbital periods experience no similar secular change within the accuracy of current determinations.[2]

It is here suggested that this linear blueshifting is instead due to a continual spontaneous increase in photon frequency which local photons normally undergo, but which until now has passed unnoticed due to the small size of the effect. For example, the rate of frequency shifting implied by the Pioneer 10 data, $2.28 \pm 0.4 \times 10^{-18}$ s$^{-1}$, would be many orders of magnitude smaller than rates detectable in the laboratory. Over a laboratory photon travel distance of 100 meters this would amount to a frequency change of one part in $10^{24}$, as compared with the U.S. Naval Observatory hydrogen maser clock system which is stable to only one part in $10^{15}$ for a one day integration time.

This alternative interpretation explains in a straightforward manner the anomalous frequency shifting of the Pioneer spacecraft signals by means of a phenomenon that has no effect on planet orbits. But, more importantly, the observed effect was predicted over a decade before the announced discovery of the Pioneer anomaly, being first mentioned in 1980 and described in later publications to be a necessary consequence of the subquantum kinetics physics methodology.[3 - 7] In 1980, the author had proposed an experiment that could test for this blueshifting effect by transponding a maser signal between two spacecraft separated by a distance of 5 AU (e.g., positioned at 1 AU and 6 AU) and the return signal being measured to determine whether its frequency had increased at the predicted average rate of $\mu \sim 1.3 \pm 0.65 \times 10^{-18}$ s$^{-1}$. The beam was to be modulated with regularly spaced pulses whose period in the return beam could be compared to the initial pulse period to determine the relative movement of the craft. In this way the Doppler component of the maser signal's frequency change would be known so that a check could be made for the presence in the beam of a nonDoppler frequency shift residual.

At that time in 1980, the author had contacted Frank Estabrook at the Jet Propulsion Laboratory (JPL), one of John Anderson's colleagues, and inquired about the possibility of conducting such a space-based experiment.[8] But at that time his group was mainly interested in using spacecraft maser signal residuals as a way of detecting gravity waves and testing the predictions of general relativity.[9 - 11] Nevertheless this discussion gave assurance that a maser signal traveling over the course of a 10 AU roundtrip journey would accrue a frequency shift large enough to be marginally measurable above the background noise that would be present in the return signal due to disturbance by the solar plasma. A description of this proposed maser signal experiment was submitted for journal publication in 1980 and was later documented in expositions of subquantum kinetics published in 1985 and 1994.[3, 4, 6]

Subquantum kinetics predicts that the magnitude of the blueshifting rate varies linearly with the magnitude of the local ambient gravitational potential. Hence, compared with the value in the Earth's environs ($1.3 \times 10^{-18}$ s$^{-1}$), for signals transponded from Earth to a spacecraft such as Pioneer 10 located at 65 AU, it predicts a slightly lower blueshifting rate of $\mu \sim 1.1 \pm 0.6 \times 10^{-18}$ s$^{-1}$. This prediction is strikingly close the observed Pioneer 10 value, being about half of the observed rate. If the Pioneer spacecraft tracking anomaly remains unexplained by more conventional phenomena, such as waste heat radiation, then the observed anomalous frequency shift may be a legitimate verification of this previously documented subquantum kinetics blueshifting prediction.

## 2. THE PHOTON BLUESHIFTING PREDICTION

Subquantum kinetics postulates that the electric and gravitational field potentials that form photons, subatomic particles, and zero-point energy fluctuations, arise as inhomogeneities in an underlying, all-pervading plenum whose constituents engage in well-defined nonequilibrium reaction-diffusion processes which are represented by a nonlinear equation system; see Ref. [4 - 6] for a full explanation. The electric field potential solutions of this equation system exhibit nonconservative as well as conservative behavior depending upon the value of the ambient gravitational potential, $\varphi_g$, relative to a critical potential value $\varphi_{gc}$; see Figure 1.[4, 6] Hence in subquantum kinetics, perfect energy conservation, photon energy remaining constant with the passage of time, is the exception rather than the rule, occurring only when this underlying reaction system operates at its critical threshold. For example, the system would operate at this threshold of marginal stability in regions of space bounding the fringes of galaxies where the ambient gravity potential approaches the critical threshold value; i.e., where $\varphi_g = \varphi_{gc}$. In regions where $\varphi_g < \varphi_{gc}$, such as in a galaxy's gravity well, *supercritical* conditions would prevail, dictating a progressive increase in photon energy with the passage of time and spontaneous photon blueshifting. In intergalactic regions of space where gravity potential attains positive values relative to the critical threshold, $\varphi_g > \varphi_{gc}$, *subcritical* conditions would prevail, dictating a progressive decrease in photon energy with the passage of time and photon redshifting. Photons traveling from distant galaxies would be subject to predominantly subcritical conditions along their flight path and hence would undergo a redshift scaling in proportion to their distance or time of travel. This "tired-light" effect model has been shown to make a good fit to cosmological redshift data on a variety of cosmology tests.[12-14] The nonconservative energy behavior predicted by subquantum kinetics does not contradict the first law of thermodynamics, since the First Law strictly applies to closed systems, whereas the nonequilibrium subquantum reactions postulated in subquantum kinetics instead constitute an *open system*.

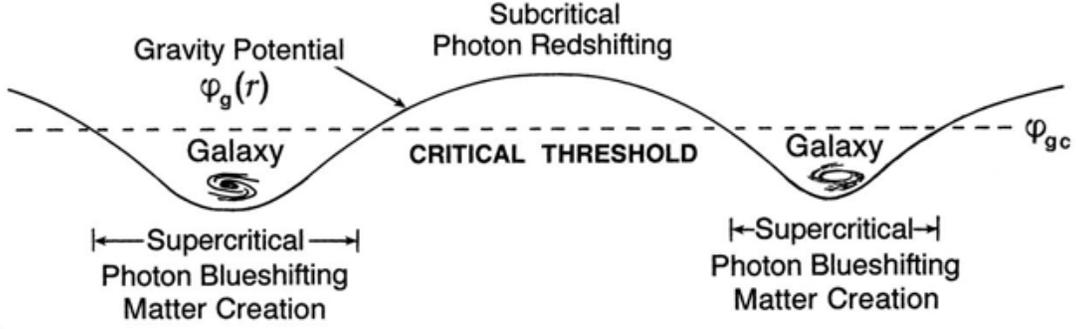

Figure 1. Hypothetical intergalactic gravity potential field and its creation of supercritical and subcritical regions.

The absolute value of the gravitational potential is important in that as a control factor it determines not only whether a photon will be in an energy gaining or energy losing mode, but also the degree to which photons will depart from perfect energy conservation, i.e., the rate at which photon energy will either increase or decrease. Subquantum kinetics represents the time dependence of photon energy by the following general relation:

$$E(t) = E_0 e^{\mu t}, \qquad (1a)$$

$$\text{where } \mu = -\alpha(\varphi_g - \varphi_{gc}). \qquad (1b)$$

$E_0$ representing the photon's initial energy, $E(t)$ its energy after time t, and $\mu$ its rate of energy change. Coefficient $\mu$ varies linearly with the value of the ambient (negative) gravitational potential $\varphi_g$ as indicated in (1b), where $\varphi_{gc}$ is the critical threshold gravity potential and $\alpha$ is a proportionality constant having units of time/area (e.g., $s/cm^2$).

In order to simplify the above equation, let us assign a value of zero to the critical threshold $\varphi_{gc}$. Then, negative gravity potential values ($\varphi_g < 0$) relative to the critical threshold zero point dictating supercritical, energy amplifying conditions and positive values ($\varphi_g > 0$) dictating subcritical, energy damping conditions, but, in addition, $\varphi_g$ determines the *rate* at which a photon's energy is expected to change over time. Alternatively, (1a) may be expressed as:

$$\nu(t) = \nu_0 e^{-\alpha \varphi_g t}, \qquad (2)$$

where $\nu_0$ is the photon's initial frequency and $\nu(t)$ its frequency at time t. This frequency shifting effect may also be expressed as:

$$\frac{d\nu}{dt} = \mu\nu = -\alpha\varphi_g \nu \qquad (3)$$

Note that this frequency shift is a new effect that is not predicted by standard physics theories and should therefore not be confused with the better known gravitational frequency shift phenomenon which arises due to the difference in ambient gravity potential between the region of photon emission and region of photon reception. The hypothesized phenomenon would occur even when there is virtually no net change of gravity potential, as in the case of maser signals transponded over a round-trip path. Also unlike the gravitational redshift, the proposed phenomenon should have no observable effect on photon emission processes, since many hours are required before a photon has accumulated a frequency shift large enough to be observed, and even then the amount is very slight. So, this effect should not be observable in stellar spectra. Over a distance of 10 kpc, photons traveling from a distant star would accumulate a blueshift of approximately 0.3 km/s, an amount that would be insignificant in comparison to shifts attributable to the conventional gravitational frequency shift phenomenon.

To provide a realistic cosmology, the reaction-diffusion system postulated as the basis of subquantum kinetics must operate very close to its threshold of marginal stability in an initially subcritical state prevailing during the era prior to the emergence of quanta. This allows the reaction system to spawn material particles out of the prevailing zero-point energy fluctuation background. Moreover by designing the reaction system so that the parameter regulating system criticality is identified with gravity potential, the reaction system is able to generate supercritical conditions in the vicinity of materialized particles and celestial bodies (i.e., in gravity wells) thereby ensuring that their forms are sustained over time. As a consequence of this cosmology, photons traveling in supercritical regions within celestial bodies and in their environs would gradually blueshift. The manner in which a photon's field potential would evolve over time is specified by (1a) above.

To take the next step and make a quantitative prediction of the rate of photon blueshifting that would be expected for a given gravity potential value, one must constrain the reaction system model with observation. Since the subquantum level is inherently unobservable, its variables being hidden from us, in order to constrain the constants in (1b) we are left to use physical observations such as bolometric luminosities for the Sun and planets. This may be done by adjusting parameter $\alpha$ and the value assumed for the galactic gravity background potential, $\varphi_{gal}$, to give a realistic value of $\mu$ as a function of $\varphi_g$ such that the resulting energy blueshifting rate yields the correct bolometric luminosities for these celestial bodies. This modeling was done in earlier discussions of the photon blueshifting effect,[3,5] and is reviewed and updated in the next section. Once this modeling was done and the subquantum kinetics photon blueshifting prediction was accordingly defined, the resulting blueshifting rate then emerged as a testable prediction, one that could later be checked through the proposed maser signal experiment. As is shown below, the Pioneer spacecraft data later confirmed this prediction.

Understandably, subquantum kinetics takes a different approach than that used in general relativity. For example, its gravitational effects are not theorized to result from a warping of space-time. Rather, effects such as gravitational attraction or repulsion, gravitational time dilation, and the bending of starlight by celestial bodies are effects that emerge as consequences of the postulated subquantum reactions.[14] The postulated subquantum reaction processes also predict that electric and gravitational potential fields should be coupled at all energies, a result that has been verified.[4, 14] Note that the proposal that gravity potential controls the rate of photon energy change, itself presupposes the existence of a link between gravitational and electric potential fields. Hence subquantum kinetics qualifies as a unified field theory.

## 3. EARLY MODELING OF THE AMPLIFICATION COEFFICIENT

Equation (3) predicts that the energy reservoir within a celestial body should spontaneously generate excess energy ("genic" energy) through this photon blueshifting effect at a rate equal to the product of the blueshifting rate coefficient $\mu$ and the reservoir's total heat capacity, $H$:

$$L_g = dE/dt = \mu H \sim -\alpha \varphi_g \bar{C} M \bar{T}, \tag{4}$$

where $H$ is given approximately as the product of the body's average specific heat $\bar{C}$, mass $M$, and average internal temperature $\bar{T}$.[4, 6] This blueshifting effect is expected to act in the same way on all photons regardless of their frequency. In this model, heat capacities $H_i$ are estimated for each body based on the body's mass, and on a reasonable estimate of its average internal temperature and specific heat. Furthermore, a planet's average internal gravity potential is modeled to be approximately $\varphi_g = 2\varphi_0 + \varphi_{sun} + \varphi_{gal}$, where $\varphi_0$ is its surface gravity potential, $\varphi_{sun}$ is the gravity potential contribution from the Sun at the planet's heliocentric distance, and $\varphi_{gal}$ is an additional background factor contributed by the Galaxy as a whole relative to the local intergalactic background potential. When variables $\alpha$ and $\varphi_{gal}$ are properly chosen, (4) is found to account for most or all of the internal energy output observed to come from the interiors of the planets Earth, Jupiter, Saturn, Uranus, and Neptune.[5, 6] Genic energy luminosities $L_g$ estimated for the Sun, planets, and Sirius B are presented in Table I along with the values for $\varphi, \bar{C}, M$, and $\bar{T}$, for comparison to observed luminosities. This table is similar to that presented in an earlier publication of this prediction, Table II of Ref. 5, but with slightly revised values for $\mu$, calculated by assuming $\alpha = 2.62 \times 10^{-32}$ s/cm² and $\varphi_{gal} = -4 \times 10^{13}$ cm²/s².

The photon blueshifting predicted for the Earth environs is so small as to be undetectable in the laboratory, but it should be large enough to be measurable over interplanetary distances. Using the modeled values for $\alpha$ and $\varphi_{gal}$, it is possible to make a testable prediction of the blueshifting rate one would expect to find in interplanetary space. That is, expressing (3) as $\mu = -\alpha \varphi_g = -\alpha (\varphi_{gal} +$

**Table I:** Modeling Parameters and Intrinsic Luminosities for the Sun, Planets and Sirius B.

| Star or Planet | M (g) | R (cm) | $\varphi_0$ (cm²/s²) | r (cm) | $\varphi_{sun}$ (cm²/s²) | $\varphi_{gal}$ (cm²/s²) | $\bar{\mu}$ (s⁻¹) | $\bar{C}$ (erg/g/°K) | $\bar{T}$ (°K) | $L_g$ (erg/s) | $L_i$ (erg/s) |
|---|---|---|---|---|---|---|---|---|---|---|---|
| Sun | 1.99 (33) | 6.96 (10) | -1.91 (15) | --- | --- | -4.0 (13) | 1.00 (-16) | 2.1 ± 0.8 (8) | 9.5 ± 3 (6) | 4.0 ± 2.5 (32) | 3.90 (33) |
| Mercury | 3.30 (26) | 2.44 (8) | -9.02 (10) | 6.00 (12) | -2.21 (13) | -4.0 (13) | 1.63 (-18) | 1.26 ± 0.5 (7) | 2 ± 1 (3) | 1.4 ± 0.9 (19) | --- |
| Venus | 4.87 (27) | 6.05 (8) | -5.37 (11) | 1.12 (13) | -1.19 (13) | -4.0 (13) | 1.39 (-18) | 1.26 ± 0.5 (7) | 2.5 ± 1 (3) | 2.1 ± 1.4 (20) | --- |
| Earth | 5.98 (27) | 6.38 (8) | -6.25 (11) | 1.50 (13) | -8.87 (12) | -4.0 (13) | 1.31 (-18) | 1.26 ± 0.5 (7) | 2.5 ± 1 (3) | 2.5 ± 1.5 (20) | 4.0 ± 0.2 (20) |
| Moon | 7.35 (25) | 1.74 (8) | -2.82 (10) | 1.50 (13) | -8.87 (12) | -4.0 (13) | 1.28 (-18) | 1.26 ± 0.5 (7) | 2 ± 1 (3) | 2.4 ± 1.5 (18) | 7.0 ± 0.5 (18) |
| Mars | 6.44 (26) | 3.39 (8) | -1.27 (11) | 2.36 (13) | -5.62 (12) | -4.0 (13) | 1.20 (-18) | 1.26 ± 0.5 (7) | 2 ± 1 (3) | 2.0 ± 1.2 (19) | 3 ± 2 (19) |
| Jupiter | 1.90 (30) | 6.92 (9) | -1.83 (13) | 8.06 (13) | -1.65 (12) | -4.0 (13) | 2.05 (-18) | 1.18 ± 0.5 (8) | 9 ± 5 (3) | 4.1 ± 2.6 (24) | 3.4 ± 0.3 (24) |
| Saturn | 5.69 (29) | 5.73 (9) | -6.62 (12) | 1.48 (14) | -8.99 (11) | -4.0 (13) | 1.42 (-18) | 8.1 ± 3.0 (7) | 6 ± 3 (3) | 3.9 ± 2.5 (23) | 8.6 ± 0.1 (23) |
| Uranus | 8.74 (28) | 2.57 (9) | -2.27 (12) | 2.97 (14) | -4.47 (11) | -4.0 (13) | 1.18 (-18) | 3.8 ± 1.5 (7) | 4 ± 2 (3) | 1.6 ± 1.0 (22) | 0.3 ± 0.4 (22) |
| Neptune | 1.03 (29) | 2.53 (9) | -2.72 (12) | 4.65 (14) | -2.85 (11) | -4.0 (13) | 1.17 (-18) | 3.6 ± 1.5 (7) | 4 ± 2 (3) | 1.7 ± 1.1 (22) | 3.3 ± 0.4 (22) |
| Pluto | 6.60 (26) | 2.90 (8) | -1.52 (11) | 4.65 (14) | -2.85 (11) | -4.0 (13) | 1.06 (-18) | 1.26 ± 0.5 (7) | 2 ± 1 (3) | 1.8 ± 0.7 (19) | --- |
| Sirius B | 2.10 (33) | 5.6 (8) | -2.5 (17) | --- | -5.0 (17) | -4.0 (13) | 1.3 (-14) | 3.0 ± 1.5 (6) | 1 ± 0.8 (6) | 8.3 ± 5.3 (31) | 1 ± 0.2 (32) |

<sup>a</sup> The numbers in brackets represent powers of 10. The values for the model parameters listed in Table I were determined as follows. For all celestial bodies considered here, the gravity potential is calculated relative to a background value of $\varphi_{gal} = -4 \times 10^{13}$ cm²/s², which includes the gravity potential contribution of the Galaxy, galaxy cluster, and supercluster. The average internal gravity potential $\varphi_g$ for the Sun is estimated to be two times its surface potential, $2\varphi_0$, plus $\varphi_{gal}$, where $\varphi_0 = -M_\odot k_g / R_\odot$, $k_g$ being the gravitational constant and $R_\odot$ being the Sun's radius. The internal gravity potentials for the planets, including the Earth and Moon, are calculated as: $\varphi_g = 2\varphi_0 + \varphi_{sun} + \varphi_{gal}$, where $\varphi_{sun} = -M_\odot k_g / r$ represents the contribution from the Sun's gravity potential field at the planet's heliocentric distance, $r$. The $\varphi_g$ values for the planets are dominated primarily by the Galactic component. In the case of Sirius B, the potential is calculated to be $\varphi_g = 2\varphi_0$.

The values for $\bar{\mu}$ are calculated as $\bar{\mu} = -\alpha \varphi_g$, with $\alpha = 2.62 \times 10^{-32}$ s/cm². The value for $\alpha$ is chosen such that the calculated genic energy luminosity for the Sun is normalized to 0.1 $L_\odot$, which should be allowable even in light of the results of the Sudbury Neutrino Observatory solar neutrino experiment.

Values adopted for the average specific heats and internal temperatures are the same as those given in reference [4] with the exception of the temperature for Sirius B. For a reasonable fit to be made to the observed bolometric luminosity ($L_i$) for Sirius B, we must choose an average temperature for its interior that is an order of magnitude lower than temperatures normally modeled for its core. This is permissible if it is assumed to have a deep convective layer capable of easily conveying heat to its surface. The $L_i$ data point for Sirius B is taken from F. Paerels, et al. *Ap. J.* **329** (1988): 849-862.

$M_\odot G / r$), where $M_\odot$ is the Sun's mass, $r$ is the maser photon's heliocentric distance in AU, and $G$ is the gravitational constant, and choosing $\alpha = 2.62 \times 10^{-32}$ s/cm² and $\varphi_{gal} = -4 \times 10^{13}$ cm²/s², the interplanetary blueshifting rate at a given distance $r$ from the Sun is given as:

$$\mu = (1.05 + \frac{0.22}{r}) \times 10^{-18}. \tag{5}$$

According to (5), a maser signal making a round-trip journey between the Earth and a spacecraft located at 65 AU would blueshift at the average rate of $\mu \sim 1.05 \times 10^{-18}$ s⁻¹, the second term in the equation making a negligible contribution. This falls close to blueshifting rates later observed for signals transponded back from Pioneer 10 and 11.

Equation (5) predicts that the rate of blueshifting should be greater at times when the maser signal transponded between the Earth and the spacecraft has a trajectory that takes it near the Sun.

Hence the blueshifting incurred over the round-trip signal path between the Earth and spacecraft would be expected to fall to a minimum when the spacecraft was at solar opposition and to rise to a sharp peak when the spacecraft approached solar conjunction, at which point the maser signal would at one point reach a minimum heliocentric distance of $r = 5 \times 10^{-4}$ AU. The modeled $\alpha$ and $\varphi$ values predict that this annual effect should cause the cumulative round-trip blueshift to vary by about ±2%. Anderson *et al.* do report an annual variation in the magnitude of the Pioneer spacecraft anomalous acceleration which is similarly strongly peaked at solar conjunction and falls to a minimum at solar opposition. However, the annual variation they observe is much larger, on the order of ±30% of the average blueshifting rate; see Figure 1 of Ref. 15 or Figure 14 of Ref. 2. They propose that this variation may be due to unexplained modeling errors in the Earth's orbital orientation or in the accuracy of the planetary ephemeris. If such is the case, until they are corrected, these modeling error uncertainties will mask the annual variation predicted by (5).

Due to the uncertainty in knowing the true value of the Galactic gravity potential in the Sun's vicinity, $\varphi_{gal}$ is used here as a modeling variable whose value, together with that of coefficient $\alpha$, is chosen so that (4) makes a best overall fit to the bolometric luminosities of the Sun and planets. In an earlier paper describing this predicted blueshifting effect,[5] these variables were instead modeled as $\alpha = 5.23 \times 10^{-32}$ s/cm$^2$ and $\varphi_{gal} = -2 \times 10^{13}$ cm$^2$/s$^2$, for their fit to observed luminosity data, which specified the blueshifting coefficient as $\mu = (1.05 + \frac{0.46}{r}) \times 10^{-18}$. The revised values reflect a fit to new luminosity data which requires a lower genic energy contribution to the Sun's bolometric luminosity. Nevertheless, at large r these two models make essentially the same prediction, deviating by <0.6% at r = 40 AU. Hence in regard to the Pioneer maser data, both models predict essentially the same average blueshifting rate for $\mu$ since this data spans heliocentric distances ranging from ~20 to 65 AU where the second term is very small. Future space based experiments carried out in the inner solar system region around 1 AU should determine whether the second term in (5) is properly modeled.

Note that the earlier published derivation of $\mu$ was done with no foreknowledge of the Pioneer Effect since the latter had not been discovered at that point. Also adjustments of this $\mu$ value made in the present paper were made solely with the intent to arrive at an adequate fit to the solar and planetary luminosity data. For example, this new fit makes a much lower genic energy prediction for the Sun, which accords with the updated lower main sequence M-L data discussed below. This new adjustment of the values of $\alpha$ and $\varphi_{gal}$ took into account no consideration of the Pioneer Effect results. Moreover, as noted above, the earlier and present model predictions for $\mu$ make the same quantitative prediction for the Pioneer data. Hence the subquantum kinetics blueshifting prediction stands as a valid *a priori* prediction.

The modeled value of $\varphi_{gal} = -4 \times 10^{13}$ cm$^2$/s$^2$ is low compared to what standard theories predict for the Galactic contribution to the local gravity potential. However, it is nevertheless consistent with observation. For example, Olling and Merrifield have modeled the Milky Way's rotation and found that the available data is best fit if the galactocentric distance for the Sun is set at $r_0 = 7.1 \pm 0.4$ kpc with a solar orbital velocity of $v_0 = 184 \pm 8$ km/s.[16-19] These values are smaller than those of other Galaxy dynamics models, which assume $r_0 = 8.5$ kpc and $v_0 = 220$ km/s. However, a smaller value for $r_0$ is corroborated by the best primary measurements which place the galactocentric distance at $7.2 \pm 0.7$ kpc.[20] Olling and Merrifield estimate a mass for the Galaxy's central bulge of $M_b \sim 8.3 \times 10^9 \, M_\odot$, based on a bulge k-band luminosity of $L_b \sim 1.5 \times 10^{10} \, L_\odot$ and a mass-luminosity ratio of 0.55.[19] For a galactocentric distance of $r_0 = 7.1$ kpc, this would contribute a gravity potential of $-5 \times 10^{13}$ cm$^2$/s$^2$ to the solar vicinity. Gerhard[21] estimates a larger mass for the galactic bulge of $1.3 \times 10^{10} \, M_\odot$ which predicts a larger contribution of $-7.8 \times 10^{13}$ cm$^2$/s$^2$. The Galaxy's disc makes a smaller contribution to the local gravitational potential. Kuijken and Gilmore have determined that the matter column density within 1.1 kpc of the galactic plane is $\Sigma_{1.1 \, kpc} = 71 \pm 6 \, M_\odot$/pc$^2$ which includes the halo contribution as well.[22] Integrating this contribution out to a distance of 7 kpc yields a gravity potential contribution to the solar vicinity of $2\pi r G \Sigma_{1.1 \, kpc} = -4.2 \pm 0.4 \times 10^{13}$ cm$^2$/s$^2$. Hence including the larger estimate for the bulge contribution, the total Galactic gravity potential contribution to the solar vicinity would be $\Delta\varphi_{gal} \sim -12 \pm 3 \times 10^{13}$ cm$^2$/s$^2$.

To obtain $\varphi_{gal}$, the Galactic gravity potential contribution $\Delta\varphi_{gal}$ must be added to the value of the intergalactic background potential which according to subquantum kinetics is positive relative to $\varphi_{gc}$, the zero critical point value; see Figure 1. An average value for the intergalactic background potential may be obtained from the cosmological redshift as: $\bar{\varphi}_{int} = H_0/\alpha$, where $H_0$ is the Hubble constant expressed in seconds$^{-1}$. As mentioned earlier, subquantum kinetics interprets the cosmological redshift as a tired-light effect, a spontaneous loss of energy specified by (1) where $\varphi_g$ is in this case positive relative to $\varphi_{gc}$. This yields an average intergalactic background potential that is approximately the same magnitude as the solar system value except opposite in sign with respect to the critical threshold zero point value. Taking $H_0 = 72 \pm 10$ km/s/Mpc and $\alpha = 2.62 \times 10^{-32}$ s/cm$^2$, we obtain $\bar{\varphi}_{int} = 8.9 \pm 1.2 \times 10^{13}$ cm$^2$/s$^2$. Adding to this $\Delta\varphi_{gal}$ would yield $\varphi_{gal} = -3 \pm 3 \times 10^{13}$ cm$^2$/s$^2$. which is close to the $-4 \times 10^{13}$ cm$^2$/s$^2$ value we are assuming.

Subquantum kinetics predicts that over great distances the gravity potential field should taper off to a plateau in intergalactic space, each galaxy's field being effective n its own gravity well locale.[4, 6]. This is similar in many respects to MOND (Modified Newtonian Dynamics) which is an observationally based approach that allows the plateau in galaxy orbital velocities to be modeled without introducing assumptions about outlying hidden mass.[23-25] In subquantum kinetics, however, this field tapering emerges as a prediction of the theory itself which requires that potential fields be slightly *nonsolenoidal*, that at large *r* the gravity potential should be less negative than would be expected from the Newtonian 1/*r* relation. Translated into more familiar terms, this would be similar to summing a galaxy's potential at a given point with the potential produced by a distributed background of "virtual antimatter" which presents a negative mass density (positive $\varphi_g$). At increasing distances from a galaxy, an increasing volume of this negative mass would be encompassed within the sphere defined by that radial distance and this would have the effect of increasingly screening the galaxy's positive mass, progressively reducing its potential from what would normally be predicted by a Newtonian $1/r$ law. Thus if this distributed background is taken to be equivalent to a negative mass density of $\rho = -10^{-30}$ g/cm$^3$, which approximates the negative of the intergalactic mass density, then a Galaxy of mass 5 X 10$^{10}$ $M_\odot$ would have its gravity field entirely screened at a galactocentric distance of 3 million light years at which point the gradient of its potential field would be zero. To satisfy MOND, a more rapid attenuation of the gravity field would be required, MOND requiring that attenuation of the gravitational force begins to be noticeable when the field's gravitational acceleration has diminished below a critical threshold of $a_0 = 10^{-8}$ cm/s$^2$.

The idea of a truncated range for gravity is consistent with the observation of Tifft that neighboring and distant galaxies do not appear to dynamically interact with one another.[26] Moreover others have argued on theoretical grounds that a range limit to gravity would avoid problems associated with space having an arbitrarily high or infinite value for its gravity potential contribution from extragalactic sources.[27] A finite range for gravity would be problematic for models that predict galaxy formation through gradual accretion of a primordial gas cloud. However, this is not a problem with nonstandard creation models such as subquantum kinetics which predict galaxy formation through continuous matter creation in a galaxy's core and gradual outward growth with later stage formation of spiral arms, a scenario supported by Hubble telescope observations of distant galaxies.[14]

## 4. ASTROPHYSICAL CONSEQUENCES OF PHOTON BLUESHIFTING

The existence of photon blueshifting and the resulting production of "genic energy" would have significant implications for astrophysics in that genic energy would make a major contribution as a stellar power source, supplementing nuclear fusion and gravitational accretion. For example, (4) predicts that red dwarfs, brown dwarfs, and planets should be powered exclusively by genic energy and should share a common mass-luminosity relation of approximately $L \propto M^{2.7 \pm 0.9}$; see ref. (5) for details. In fact, the jovian planets are seen to lie along the lower main sequence stellar mass-luminosity relation which, using the data of Harris *et al.*,[22] is of the form $L \propto M^{2.76 \pm 0.15}$; see Fig. 2.[5, 6] The brown dwarfs LP 944-20, G 196-3B, and GL 229B [29-32] also are found to lie close to this line confirming earlier predictions.[4 - 6] These correspondences provide strong confirmation for the subquantum kinetics blueshifting prediction since on standard theory they must be passed off as pure coincidence.

Between 0.4 $M_\odot$ and 0.6 $M_\odot$ (L ≈ 0.025 - 0.05 $L_\odot$), the mass-luminosity relation becomes somewhat scattered and less well defined. It is in this transition region where the upper main sequence mass-luminosity relation begins, the M-L relation changing to a steeper slope of $L \propto M^4$; (dashed line in Fig. 2). This transition would signify the point where nuclear burning begins to make a noticeable contribution, supplementing genic energy production. When projected to lower masses, the upper main sequence M-L relation is found to intersect the lower main sequence M-L relation at around 0.4 $M_\odot$. But the lower main sequence relation data trend does not entirely disappear until a somewhat higher mass is reached, possibly around 0.6 to 0.7 $M_\odot$. The luminosity gap evident between the lower and upper main sequence relations would indicate the additional contribution provided by nuclear fusion, nuclear burning providing a progressively larger share of a star's total radiated energy as stellar mass increases.

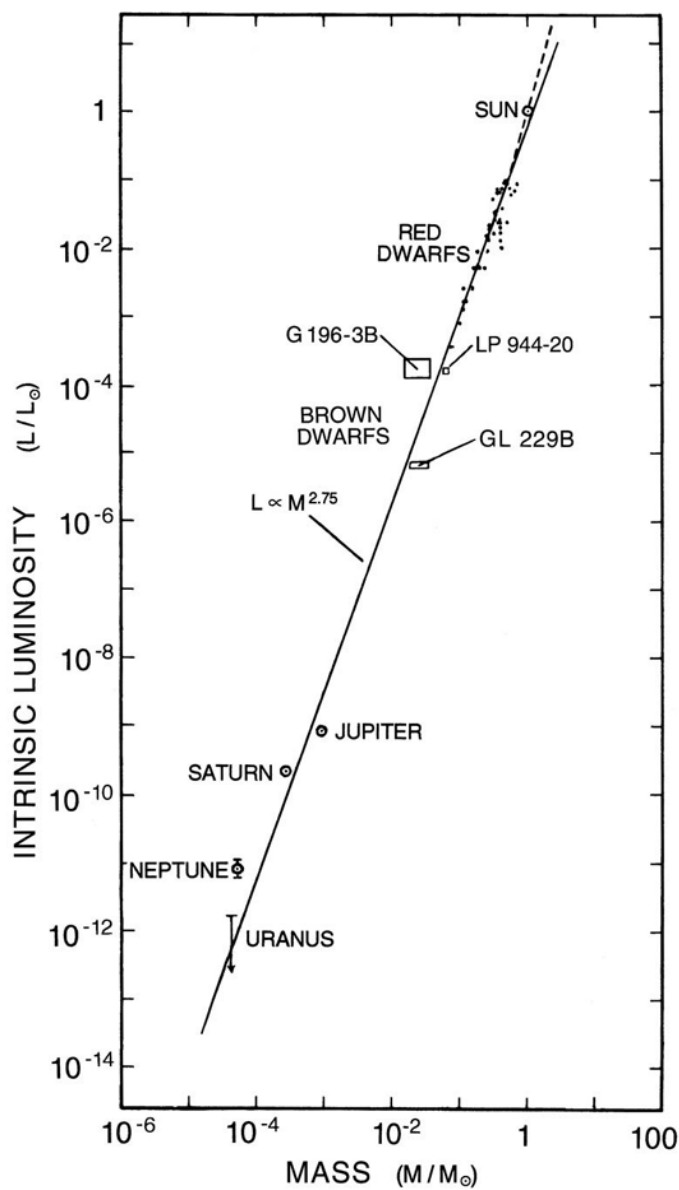

Figure 2. The mass-luminosity coordinates of the jovian planets shown
in relation to the lower main sequence stellar mass-luminosity relation.
The M-L coordinates for the recently discovered brown dwarfs (LP 944-20,
G 196-3B, and GL 229B) also are found to lie close to this line.

In this mass range of 0.2 to 0.5 $M_\odot$, where the lower main sequence transitions to the upper main sequence, the lower log M - log L relation gradually bends downward to a slightly lower slope.

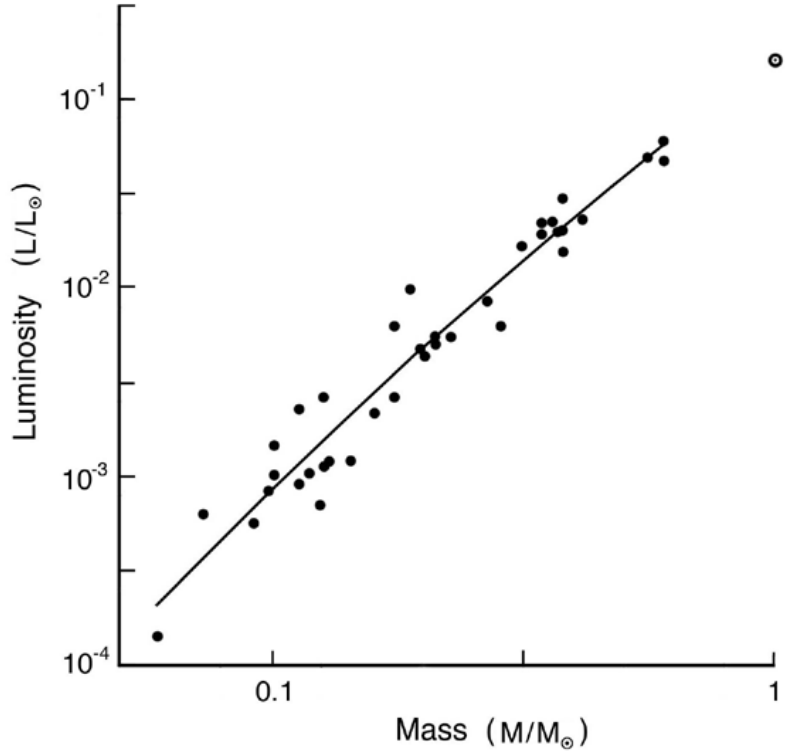

Figure 3. A nonlinear curve fit to red dwarf mass-luminosity data. The circle in the upper right hand corner is the indicated value when the regression is projected up to 1 $M_\odot$.

This may be seen in Figure 3, which plots a nonlinear regression curve fit made to log M and log L data for red dwarfs in the mass range 0.06 $M_\odot$ through 0.6 $M_\odot$. The plotted values are listed in Table II. In cases where the cited sources list visual magnitudes instead of luminosity values, bolometric luminosities are calculated using the models of Baraffe, et al. by knowing the spectral classifications of the respective stars.[33] This downward bend, or "plateau," has also previously been noted by Henry et al. who have attributed the effect partly to a deepening of a star's convective region as stellar mass increases.[34]

When extrapolated upward to 1 $M_\odot$, the resulting regression line projects a bolometric luminosity of $0.16 \pm 0.07$ $L_\odot$ for the Sun. Subquantum kinetics, then, predicts that around 16% of the Sun's energy should be of non-nuclear genic origin with the remaining majority being supplied by nuclear fusion.[35] While the Sudbury Neutrino Observatory solar neutrino experiment results have resolved the mystery as to why other solar neutrino experiments had previously reported low solar neutrino fluxes, the proposal that fusion may provide only $84 \pm 7$ % of the Sun's energy is nevertheless tolerable in view of the range of uncertainty of stellar fusion models.

If the existence of the photon blueshifting phenomenon is acknowledged, all stellar evolution models will need to be revised. As described in previous publications, genic energy is also able to account for high-luminosity white dwarfs and X-ray stars and can explain how certain quasars are able to power themselves in the absence of nearby gas and dust, a recognized problem for black hole theory.[4-6, 14] Moreover genic energy can account for phenomena such as stellar pulsation, novae, and supernovae. That is, since photon blueshifting can feed back to increase a star's internal temperature and heat capacity which in turn increases the rate at which energy is produced, this inherent nonlinearity can ultimately lead to instability.

**Table II:** Masses and Luminosities of Low Mass Stars

| Star | $\log \frac{M}{M_\odot}$ | $\log \frac{L}{L_\odot}$ | Ref. | Star | $\log \frac{M}{M_\odot}$ | $\log \frac{L}{L_\odot}$ | Ref. |
|---|---|---|---|---|---|---|---|
| GL 22 A | -0.44 | -1.64 | [a] | GL 747A | -0.67 | -2.25 | [b] |
| GL 22 C | -0.89 | -2.57 | [a] | GL 747B | -0.70 | -2.32 | [b] |
| GL 65 A | -0.99 | -2.99 | [b] | GL 748A | -0.42 | -1.80 | [c] |
| GL 65 B | -1.00 | -3.07 | [b] | GL 748B | -0.72 | -2.00 | [c] |
| GL 166 C | -0.75 | -2.20 | [a] | GL 791.2B | -0.90 | -3.15 | [b] |
| GL 234 A | -0.69 | -2.36 | [b] | GL831 A | -0.54 | -2.20 | [b] |
| GL 234 B | -0.99 | -3.08 | [b] | GL 831 B | -0.79 | -2.66 | [b] |
| GL 473 A | -0.84 | -2.91 | [b] | GL 860 A | -0.57 | -2.07 | [b] |
| GL 473 B | -0.88 | -2.91 | [b] | GL 860 B | -0.76 | -2.40 | [b] |
| GL 570 B | -0.25 | -1.30 | [b] | GL 866 A | -0.92 | -2.98 | [b] |
| GL 570 C | -0.42 | -1.52 | [b] | GL 866 B | -0.94 | -3.03 | [b] |
| GL 623 A | -0.46 | -1.71 | [a] | GL 866 C | -1.03 | -3.24 | [b] |
| GL 623 B | -0.94 | -2.64 | [a] | YY Gem A | -0.22 | -1.22 | [b] |
| GL 644 A | -0.38 | -1.63 | [b] | YY Gem B | -0.22 | -1.32 | [b] |
| GL 644 Ba | -0.46 | -1.65 | [b] | GJ 1245 A | -0.89 | -2.94 | [a] |
| GL 644 Bb | -0.50 | -1.77 | [b] | GJ 1245 C | -1.13 | -3.44 | [a] |
| GL 661 A | -0.42 | -1.69 | [b] | CM Dra a | -0.64 | -2.25 | [b] |
| GL 661 B | -0.43 | -1.7 | [b] | CM Dra b | -0.67 | -2.29 | [b] |
|  |  |  |  | LP 944-20 | -1.22 | -3.80 | [d] |

[a] T. J. Henry, O. G. Franz, L. H. Wasserman, G. F. Benedict, P. J. Shelus, P. A. Ianna, J. D. Kirkpatrick, and D. W. McCarthy, Jr., *Ap. J.* **512**, 864 (1999).

[b] X. Delfosse, T. Forveille, D. Ségransan, J.-L. Beuzit, S. Udry, C. Perrier, and M. Mayor, *Astron. Astrophys.* **364**, 217 (2000).

[c] G. F. Benedict, B. E. McArthur, O. G. Franz, L. H. Wasserman, T. J. Henry, T. Takato, I. V. Strateva, J. L. Crawford, P. A. Ianna, D. W. McCarthy, et al., *A. J.* **121,** 1607 (2001).

[d] C. G. Tinney, *MNRAS* **296**L, 42 (1998); C. G. Tinney and I. N. Reid, *MNRAS* **301**, 1031 (1998).

In an earlier 1985 paper, the author pointed out that the photon blueshifting rate estimated for photons traveling in the solar system was approximately the same magnitude as the energy attenuation rate needed to account for the cosmological redshift, but of opposite sign;[4, 6] and more recently, discoverers of the Pioneer Effect have similarly noted that the blueshifting anomaly has a magnitude comparable to $H_0$.[2] In fact, the revised estimate of the Pioneer data blueshifting rate discussed below ($2.28 \pm 0.4 \times 10^{-18}$ s$^{-1}$) is exactly equal in magnitude to the negative Hubble constant value, the subquantum kinetics prediction being about half this amount. In the context of subquantum kinetics, this correspondence between the predicted Pioneer maser signal blueshifting rate and the cosmological redshifting rate would be an indication that the value of the gravitational potential for regions of space within galaxies does not depart far below the critical threshold, just as the intergalactic value of the gravitational potential does not depart far above the critical threshold. This is consistent with the requirement in subquantum kinetics that the underlying subquantum reaction-diffusion system operates very close to the threshold of marginal stability, and that it had resided predominantly in a subcritical state during the period preceding the emergence of matter in supercritical regions. If the subquantum reactions had not initially operated near its critical threshold, they would have had little chance of spawning material particles out of the zero-point energy fluctuation background.

## 5. ADJUSTING THE PIONEER BLUESHIFTING ANOMALY DATA FOR UNMODELED THERMAL EFFECTS

As mentioned above, the magnitude of the Pioneer anomalous acceleration effect reported by Anderson *et al*. should be reduced to 82% of its value to take into account the propulsive effects due to waste heat radiated from the spacecraft. For example, Scheffer, Katz, and Murphy have suggested that a portion of the apparent anomalous acceleration acting on the Pioneer spacecraft may be due to thermal radiation striking the anti-solar side of its antenna coming from the passive radiators used to cool the spacecraft's electronics, from its RTG power unit, and from various other sources.[36 - 38] Scheffer's model predicts that the thrust from these thermal sources should have declined by 11.8% from "Period I" (~10/1988) through "Period III" (~7/1995) due to a decline in available spacecraft power and changes in the types of experiments being carried out. Instead, a much smaller rate of decrease in acceleration is seen. The SIGMA WLS values computed for the anomalous Pioneer 10 acceleration, listed in Table I of Ref. [2], indicate a decrease of 2% from Period I to Period III, whereas the CHASMP WLS values indicate a decrease of 4.1% over this same time period. Averaging these two values, it appears that the spacecraft's acceleration decreased by about 3.05 ± 1% over this period.[39] Consequently, if this 3% decrease is entirely due to the above unmodeled heating effects, these effects would be responsible for at most $(3.05 \pm 1\%)/11.8\% = 26 \pm 9\%$ of the anomalous "acceleration." Although (5) indicates there should be some degree of decrease in $\mu$ as the spacecraft's heliocentric distance progressively increases, this decline would be very slight, only 0.15% during the above mentioned interval.

Anderson *et al*., however, maintain that such unmodeled thermal effects account for a much smaller percentage of the anomalous apparent acceleration,[2,40-42] and include a bias correction of only −0.55 X $10^{-8}$ cm/s$^2$ to take into account the propulsive effect of heat radiated from the RTG unit; see Table II of Ref. [2]. They find no variation in Pioneer 10's computed acceleration over a certain period from mid 1992 to mid 1998 and offer this as evidence that the unmodeled thermal effects play a minor role. However, since their trend line was calculated from data that included the ±30% annual variation described above, their regression line slope will very much depend on which dates they pick to begin and end their trend-line average. Consequently, a 3% secular decline may still be present in their data yet be masked by the much larger irregular annual variation. Hence a negative conclusion in regard to a decline in spacecraft acceleration is unwarranted.

The above discussion of Scheffer's thermal effects model, suggests that the anomalous acceleration value of $8.7 \pm 1.3 \times 10^{-8}$ cm/s² reported by Anderson, *et al.* should be reduced by the additional amount of $1.86 \pm 1.3 \times 10^{-8}$ cm/s², i.e., $0.26 \times (8.7 + 0.55) \times 10^{-8} - 0.55 \times 10^{-8}$ cm/s², to give an unexplained residual of $6.85 \pm 1.3 \times 10^{-8}$ cm/s². Expressed in terms of photon blueshifting, this residual would amount to $2.28 \pm 0.4 \times 10^{-18}$ s⁻¹, which comes strikingly close to the blueshifting rate predicted by subquantum kinetics.

## 6. GALILEO RANGING MEASUREMENTS.

If the unexplained frequency shift phenomenon is a Doppler shift produced by an anomalous force acting on the spacecraft, as Anderson *et al.* have suggested, then one should observe a corresponding change in the spacecraft's range relative to the Earth. That is, the spacecraft should be found to be closer than would otherwise be expected. If the frequency shift instead arises spontaneously as a nonDoppler blueshift affecting only the transponded maser signal, as subquantum kinetics predicts, then one should expect to find no corresponding change in spacecraft range. Unfortunately, the Pioneer 10 and 11 spacecraft were not outfitted for ranging measurements. So Anderson *et al.* attempted to check for the presence of ranging anomalies by comparing the ranging and "Doppler shift" data received from the Galileo spacecraft. Ranging measurements were made by modulating the outgoing maser beam with markers and observing the time taken for a given marker to be transponded back to Earth. The Doppler shift measurements were made by comparing the frequencies of the outgoing and incoming maser beams. Anderson *et al.* claim that the marker timing measurements showed a decrease in spacecraft range approximating the amount that would be expected if the anomalous blueshifting they observe in their Doppler shift data were due to unmodeled forces acting on the spacecraft.[2] Consequently, they concluded that the data did not favor the alternative explanation of spontaneous frequency blueshifting, which they alternatively construe as the "time acceleration" hypothesis.

However, there are reasons for doubting their conclusion. For one thing, the data they used was recorded relatively close to the Sun (1 to 5 AU) while the Galileo spacecraft was traveling from near the Earth to Jupiter, a region where solar radiation pressure and solar plasma effects dominate the measurements. By comparison, Anderson *et al.* began studying the Doppler signal data from the Pioneer 10 and 11 spacecraft only after these spacecraft had surpassed a distance of 20 AU, at which point solar radiation pressure acceleration had decreased to $< 5 \times 10^{-8}$ cm/s². The same measure of caution should be applied to the Galileo ranging and Doppler measurements, raising the question as to the reliability of that data. In fact, Anderson *et al.* state that the solution they obtained for the Galileo spacecraft Doppler measurement data was so highly correlated with solar pressure and so complicated by mid-course orbital maneuvers that a standard null result could not be ruled out. They also acknowledge that the signal-to-noise ratio on the incoming range signal was small, requiring long integration times during which time the range of the spacecraft was constantly changing at a rate of ~6 km/s. Additional uncertainty was introduced because the identification of the code for the appropriate received ranging signal was inferred, sometimes with great difficulty, from the orbit determination programs.

Moreover, as mentioned above, Anderson *et al*. question the accuracy of the solar system modeling programs they used, speculating that modeling errors in these programs may be the cause of the large unexplained annual variation observed in residual acceleration. In the case of Pioneer 10 and 11, as the spacecraft progressed from 40 AU to 70 AU, the maser signal path length varied by a small percentage, about ±2%, due to distance variation arising from the movement of the Earth about the Sun (±1 AU vs. 55 AU), yet their data charted a large annual variation of ±25%. By comparison, Galileo's measurements were made 20 times closer to the Sun in the vicinity of 1 - 5 AU where the annual variation of the maser signal path length, relative to the total distance to the spacecraft, was at least 25 times greater. In this near solar environment, the variable annual component of the acceleration would have substantially exceeded the nonvarying component making ranging determinations even more uncertain. Furthermore, when a maser signal passes through the solar plasma, ranging measurements will experience a time delay, making the calculated spacecraft distance abnormally large, whereas the transponded return signal used in making the Doppler frequency comparison will experience a phase advance, making the calculated inferred spacecraft distance abnormally small.

Although ranging data was also available from the Ulysses spacecraft, those measurements were considered even less reliable than those from Galileo. In summary, the measurements that Anderson *et al*. used to rule out the spontaneous photon blueshifting effect are, by their own admission, fraught with difficulty. Hence their range change conclusion should be viewed with caution.

If the Pioneer Effect is in fact due to a spontaneous change in photon energy, as subquantum kinetics predicts and is not due to an anomalous force acting on the spacecraft, then it will be necessary to adjust for this effect in Doppler tracking of spacecraft. For example, consider the anomalous force of $8.7 \times 10^{-8}$ cm/s$^2$ which Anderson et al. suggest acts on the Pioneer 10 spacecraft. If this is imagined to be real and to have acted on the spacecraft for a full 30 years, then navigators will place the craft's position 390,000 km closer to the Sun than if the force were believed to be absent and the blueshifting due to a nonDoppler blueshifting effect. This is about the distance from the Earth to the Moon, and could result in quite a large error if the nature of the blueshifting is misjudged.

## 7. FUTURE EXPERIMENTS AND CONCLUSION

A definitive resolution of the issue of whether Doppler and ranging anomalies really exist may have to await a future out-of-solar system spacecraft mission such as the Pluto/Kuiper mission scheduled for launch around 2006 and which includes very sophisticated tracking technology. In addition, as a result of efforts by Turyshev et al.,[43] the European Space Agency has suggested that a new spacecraft experiment be launched into the outer solar system around 2015 to exclusively test the Pioneer Effect. The experiment would be designed to characterize the properties of the anomaly to an accuracy of at least two orders of magnitude below the anomaly's size.

Another possibility would be to make modifications to the Laser Interferometer Space Antenna (LISA) so that it could detect the predicted effect. LISA, which is due to be launched in 2010, involves three spacecraft placed in the Earth's solar orbit separated from one another by 5 million kilometers to form three legs of a triangle with a phase-locked laser beam along each leg.[44] Although the experiment is planned to be carried out fairly close to the Sun (~1 AU), solar plasma has little effect on laser beam transmission, hence should not pose a problem. LISA will be able to detect accelerations of as little as $10^{-13}$ cm/s$^2$ (equivalent to a frequency shift rate of ~$10^{-23}$ s$^{-1}$), which is about $10^5$ times smaller than the predicted effect. However, the experiment is currently being designed primarily for the purpose of detecting low frequency gravity waves in the milli Hz frequency range and as such is not focused on detecting a "DC" effect similar to that produced by the Pioneer anomaly, whether it be a constant accelerating force or a constant nonDoppler photon blueshifting effect.[45]

However, it should be possible to check for the Pioneer effect. At 1 AU from the Sun, the predicted genic energy effect is expected to blueshift each laser beam at the rate of $1.27 \pm 0.7 \times 10^{-18}$ s$^{-1}$, making the beam accumulate a frequency shift of $4.2 \times 10^{-17}$ by the end of its $10^7$ km roundtrip. For a beam wavelength of 1 μm, (f = $3 \times 10^{14}$ Hz), this would amount to a frequency difference of $\Delta f = 0.0127$ Hz. The resulting interference pattern would move through one fringe cycle every $1/\Delta f = 79$ seconds toward the lower frequency reference laser. This effect would be the same for each leg of the triangle regardless of each beam's orientation. LISA will be using photodiode detectors having an interferometer fringe resolution of $4 \times 10^{-5}$ $\lambda/\sqrt{Hz}$ which would allow this net fringe drift to be easily detected.

A nonDoppler frequency shift effect may be distinguished from Doppler shifts arising from relative drifting of the spacecraft by modulating the laser beam with regularly spaced AM pulses or polarization shift markers. Any change in the period of these time markers would indicate exclusively the Doppler component of the frequency shift since the marker period would be relatively immune to any change in laser frequency. If the laser transponder is instructed to return a signal that is offset from the original laser frequency by twice the Doppler frequency shift, this would effectively null out the Doppler component when the transponded signal is received back, and would leave the photon blueshifting effect to appear in the data coming from each leg as the primary fringe drift residual. If the Pioneer effect were instead due to an anomalous force pulling spacecraft toward the Sun, it should accelerate all three LISA spacecraft at approximately the same rate and hence would be expected to yield no net frequency shift effect.

It is hoped that the necessary design modifications will be made to LISA to allow a test to be made of this important physical effect. As a follow-up to LISA, a similar experiment could later be conducted closer to the Sun where the ambient gravity potential is more negative and where subquantum kinetics predicts that higher blueshifting rates should be observed. Until these experiments are carried out, the subquantum kinetics blueshifting prediction should provide a useful alternative to consider in interpreting the Pioneer 10 and 11 tracking anomaly.